\documentstyle[epsfig,mathptm]{mn}
\newif\ifAMStwofonts

\ifoldfss
  \ifCUPmtlplainloaded \else
    \NewTextAlphabet{textbfit} {cmbxti10} {}
    \NewTextAlphabet{textbfss} {cmssbx10} {}
    \NewMathAlphabet{mathbfit} {cmbxti10} {} 
    \NewMathAlphabet{mathbfss} {cmssbx10} {} 
  \fi
  \ifAMStwofonts
    \ifCUPmtlplainloaded \else
      \NewSymbolFont{upmath} {eurm10}
      \NewSymbolFont{AMSa} {msam10}
      \NewMathSymbol{\upi}     {0}{upmath}{19}
      \NewMathSymbol{\umu}     {0}{upmath}{16}
      \NewMathSymbol{\upartial}{0}{upmath}{40}
      \NewMathSymbol{\leqslant}{3}{AMSa}{36}
      \NewMathSymbol{\geqslant}{3}{AMSa}{3E}

    \fi
  \fi
\fi 

\ifnfssone
  \newmathalphabet{\mathit}
  \addtoversion{normal}{\mathit}{cmr}{m}{it}
  \addtoversion{bold}{\mathit}{cmr}{bx}{it}
  \newmathalphabet{\mathbfit} 
  \addtoversion{normal}{\mathbfit}{cmr}{bx}{it}
  \addtoversion{bold}{\mathbfit}{cmr}{bx}{it}
  \newmathalphabet{\mathbfss} 
  \addtoversion{normal}{\mathbfss}{cmss}{bx}{n}
  \addtoversion{bold}{\mathbfss}{cmss}{bx}{n}
  \ifAMStwofonts
    \ifCUPmtlplainloaded \else
      \UseAMStwoboldmath
      \makeatletter
      \new@mathgroup\upmath@group
      \define@mathgroup\mv@normal\upmath@group{eur}{m}{n}
      \define@mathgroup\mv@bold\upmath@group{eur}{b}{n}
      \edef\UPM{\hexnumber\upmath@group}
      \new@mathgroup\amsa@group
      \define@mathgroup\mv@normal\amsa@group{msa}{m}{n}
      \define@mathgroup\mv@bold\amsa@group{msa}{m}{n}
      \edef\AMSa{\hexnumber\amsa@group}
      \makeatother
      \mathchardef\upi="0\UPM19
      \mathchardef\umu="0\UPM16
      \mathchardef\upartial="0\UPM40
      \mathchardef\leqslant="3\AMSa36
      \mathchardef\geqslant="3\AMSa3E
    \fi
  \fi
\fi 

\ifnfsstwo
  \DeclareMathAlphabet{\mathbfit}{OT1}{cmr}{bx}{it}
  \DeclareMathAlphabet{\mathbfss}{OT1}{cmss}{bx}{n}
  \ifAMStwofonts
    \ifCUPmtlplainloaded \else
      \DeclareSymbolFont{UPM}{U}{eur}{m}{n}
      \SetSymbolFont{UPM}{bold}{U}{eur}{b}{n}
      \DeclareSymbolFont{AMSa}{U}{msa}{m}{n}
      \DeclareMathSymbol{\upi}{0}{UPM}{"19}
      \DeclareMathSymbol{\umu}{0}{UPM}{"16}
      \DeclareMathSymbol{\upartial}{0}{UPM}{"40}
      \DeclareMathSymbol{\leqslant}{3}{AMSa}{"36}
      \DeclareMathSymbol{\geqslant}{3}{AMSa}{"3E}
    \fi
  \fi
\fi 

\ifCUPmtlplainloaded \else
  \ifAMStwofonts \else 
    \def\upi{\pi}
    \def\umu{\mu}
    \def\upartial{\partial}
  \fi
\fi

\def\tfrac#1#2{{\textstyle\frac{#1}{#2}}}
\def\vect#1{{\mathbfit{#1}}}
\title
[Combining cosmological datasets]
{Combining cosmological datasets: hyperparameters and Bayesian
evidence}
\author[M.P.~Hobson, S.L.~Bridle and O.~Lahav]
{M.P.~Hobson$^1$, S.L.~Bridle$^2$ and O.~Lahav$^2$\\ 
$^1$Astrophysics Group, Cavendish Laboratory, Madingley Road, 
Cambridge, CB3 0HE, UK \\
$^2$Institute of Astronomy, Madingley Road, Cambridge, CB3 0HA, UK}
\date{Accepted ---. Received ---; in original form 15 March 2002}
\pagerange{\pageref{firstpage}--\pageref{lastpage}}
\pubyear{2002}

\begin{document}
\maketitle
\label{firstpage}

\begin{abstract}
A method is presented for performing joint analyses of cosmological
datasets, in which the weight assigned to each dataset is determined
directly by it own statistical properties. The weights are considered
in a Bayesian context as a set of hyperparameters, which are then
marginalised over in order to recover the posterior distribution as a
function only of the cosmological parameters of interest. In the case
of a Gaussian likelihood function, this marginalisation may be
performed analytically.  Calculation of the Bayesian evidence for the
data, with and without the introduction of hyperparameters, enables a
direct determination of whether the data warrant the introduction of
weights into the analysis; this generalises the standard likelihood
ratio approach to model comparison.  The method is illustrated by
application to the classic toy problem of fitting a straight line to a
set of data. A cosmological illustration of the technique is also
presented, in which the latest measurements of the cosmic microwave
background power spectrum are used to infer constraints on
cosmological parameters.

\end{abstract}


\section{Introduction}
\label{intro}

It is now common practice in cosmology to estimate the values
of cosmological parameters by a joint analysis of a number of
different datasets. The standard technique for performing such an
analysis is to assume that the datasets are statistically independent
and so take the joint likelihood function for the parameters to be
given simply by the product of the individual likelihood
functions for each separate dataset. The joint likelihood function
can then be used in the standard way to determine the optimal values
of the parameters and their associated errors.

As discussed by Lahav et al. (2000; hereinafter Paper I), however,
there exists some freedom in the relative `weight' that may be 
given to each dataset in the analysis (see also Godwin \& Lynden-Bell 1987;
Press 1996). The assignment of weights
often occurs when two or more of the datasets are inconsistent, and
is usually made in a somewhat ad-hoc manner. Typically some 
datasets are excluded from the analysis, and hence given a weight
of zero, while the remainder are analysed jointly with equal
weights. Despite its widespread use, this procedure has many
unsatisfactory features, not least of which is the subjectivity associated with
the choice of which datasets to include, and which to discard.
As advocated in Paper I, a more objective procedure for assigning
weights to the datasets is provided by the introduction of
{\em hyperparameters}. This device allows the statistical properties
of the data themselves to determine the weight given to each dataset
in the analysis. 

In Paper I, a method was presented for introducing hyperparameters
into the analysis of datasets for which each likelihood function had
a particular simple form. This technique was then applied to the
estimation of the Hubble parameter $h$ from several sets of
observations of the power spectrum of cosmic microwave background (CMB)
anisotropies. It was shown that apparently discrepant datasets
could be analysed jointly to provide a consistent estimate of $h$,
together with an associated error. This approach has also recently
been applied to the joint analysis of the baryon mass fraction in
clusters and cepheid-calibrated distances (Erdogdu, Ettori \& Lahav 2002).

In this paper, we extend the 
work of Paper I to accommodate more
general situations. In section~\ref{bayesevidence}, we review the
standard approach to Bayesian parameter estimation and discuss the
role of the Bayesian evidence in model selection. 
In section~\ref{hyper}, we present a general 
account of hyperparameters and their use in joint estimation of
parameters. We also discuss how to use the Bayesian evidence 
to decide whether or not the data support the inclusion of
hyperparameters in the first instance. In section~\ref{weighting}, we consider
the use of hyperparameters in the weighting of datasets and, in
section~\ref{gausslike}, we discuss the common special case in which the
likelihood function for each dataset is Gaussian. 
In section~\ref{toy}, we illustrate
the useful properties of the hyperparameters approach by applying
the technique to the toy problem of fitting a straight-line to
a number of datasets. A cosmological application of the
hyperparameters approach is discussed in section~\ref{cosmoapp}, in
which we perform a joint estimate of the physical baryon density
$\Omega_bh^2$ and the scalar spectral index $n$ using the most recent
sets of observations of the CMB power spectrum. Finally, our
conclusions are presented in section~\ref{concs}.

\section{Bayes' theorem and evidence}
\label{bayesevidence}

Suppose the totality of our data is represented by the data vector 
$\vect{D}$ and we are interested in estimating the values of the
parameters $\btheta$ in some underlying model of the data. The
standard approach to this problem is to use Bayes' theorem
\begin{equation}
\Pr(\btheta|\vect{D})=\frac{\Pr(\vect{D}|\btheta)\Pr(\btheta)}
{\Pr(\vect{D})},
\label{bayes}
\end{equation}
which gives the posterior distribution $\Pr(\btheta|\vect{D})$ 
in terms of the likelihood $\Pr(\vect{D}|\btheta)$, the prior 
$\Pr(\btheta)$ and the evidence $\Pr(\vect{D})$    
(which is also often called the marginalised likelihood). 

\subsection{Parameter estimation}

For the purpose of estimating parameters, one usually
ignores the normalisation factor $\Pr(\vect{D})$ in Bayes' theorem,
since it does not depend on the parameters $\btheta$.
Thus, one instead works with the `unnormalised posterior'
\begin{equation}
\overline{\Pr}(\btheta|\vect{D}) \equiv
\Pr(\vect{D}|\btheta)\Pr(\btheta),
\label{pbardef}
\end{equation}
where we have written $\overline{\Pr}$ to denote the fact that 
the `probability distribution' on the left-hand side is
not normalised to unit volume. In fact, it is also common to omit
normalising factors, that do not depend on the parameters $\btheta$,
from the likelihood and the prior. As we shall see below, however, 
if one wishes to calculate the Bayesian evidence for a particular
model, the likelihood and the prior must be properly normalised
such that $\int\Pr(\btheta)\,d\btheta=1$ and
$\int\Pr(\vect{D}|\btheta)\,d\vect{D}=1$. We will therefore 
assume here that the necessary normalising factors have been retained.

Strictly speaking, the {\em entire} (unnormalised) 
posterior is the Bayesian inference
of the parameters values. Unfortunately, if the dimension $M$
of the parameter space is large, it is often numerically unfeasible to
calculate $\overline{\Pr}(\btheta|\vect{D})$ on some $M$-dimensional
hypercube. Thus, particularly in large problems,  it is usual to 
present one's results in terms of the `best' estimates
$\hat{\btheta}$, which maximise the (unnormalised) posterior, 
together with some associated errors. These errors are usually
quoted in terms of the estimated covariance matrix
\begin{equation}
\vect{C} = \left[-\left.\nabla\nabla\ln\overline{\Pr}(\btheta|\vect{D})
\right|_{\btheta=\hat{\btheta}}\right]^{-1}.
\label{covmatdef}
\end{equation}
or as confidence limits on each parameter $a_i$ $(i=1,\ldots,M)$, 
obtained from the one-dimensional marginalised (unnormalised) 
posterior distributions
\begin{equation}
\overline{\Pr}(a_i|\vect{D}) 
= \int \overline{\Pr}(\btheta|\vect{D})\,d\check{\btheta},
\end{equation}
where $d\check{\btheta} = da_1\cdots d\check{a}_i\cdots da_M$ denotes
that the integration is performed over all other parameters $a_j$ 
$(j \neq i)$.

The estimates $\hat{\btheta}$ are most often obtained by 
an iterative numerical minimisation algorithm. 
Indeed, standard numerical algorithms are generally
able to locate a local (and sometimes global) maximum of this 
function even in a space of large dimensionality. Similarly, the
covariance matrix of the errors can be found straightforwardly by
first numerically evaluating the Hessian matrix 
$\nabla\nabla\overline{\Pr}(\btheta|\vect{D})$ at the peak 
$\btheta=\hat{\btheta}$, and then calculating (minus) its inverse.

\subsection{Bayesian evidence and model selection}
\label{evidmodel}

The standard technique outlined above produces inferences of
the parameter values for a given model of the data, but it does not
provide a mechanism for deciding which one of a set of alternative
models is most suitable for describing the data. This problem may be
addressed, however, using the Bayesian evidence $\Pr(\vect{D})$. 
Very readable introductions to this topic are given by Bishop (1995)
and Sivia (1996).

Although the evidence term is usually ignored in the 
process of parameter estimation, it is central to selecting 
between different models for the data. For illustration, let us
suppose we have two alternative models (or hypotheses) for 
the data $\vect{D}$; these hypotheses are 
traditionally denoted by $H_0$ and $H_1$. Let us assume further that
the model $H_0$ is characterised by the parameter set $\btheta$,
whereas $H_1$ is described by the set of parameters $\bphi$. 
For the model $H_0$, the probability density for an observed 
data vector $\vect{D}$ is given by
\begin{equation}
\Pr(\vect{D}|H_0) 
=  \int \Pr(\vect{D}|\btheta) \Pr(\btheta)\,d\btheta,
\label{eviddef0}
\end{equation}
where, on the left-hand side, we have made explicit the 
conditioning on $H_0$. Similarly, for the model $H_1$,
\begin{equation}
\Pr(\vect{D}|H_1) 
= \int \Pr(\vect{D}|\bphi) \Pr(\bphi)\,d\bphi.
\label{eviddef1}
\end{equation}
In either case, we see that the evidence is given by the average of
the likelihood function with respect to the prior. Thus, a model will
have a larger evidence if more of its allowed parameter space
is likely, given the data. Conversely, a model will have a small
evidence if there exist large areas of the allowed parameter space 
with low values of the likelihood, even if the likelihood function is strongly
peaked and the corresponding model predictions 
agree closely with the data. 
Hence the value of the evidence naturally incorporates
the spirit of Ockham's razor: a simpler theory, having a more
compact parameter space, will generally have a larger evidence than
a more complicated theory, unless the latter is 
significantly better at explaining the data.

The question of which is the models $H_0$ and $H_1$ is prefered 
is thus answered simply by comparing the relative
values of the evidences 
$\Pr(\vect{D}|H_0)$ and $\Pr(\vect{D}|H_1)$. The hypothesis
having the larger evidence is the one that should be
accepted. Although not widely used in cosmology, the idea of
model selection using evidence ratios has been considered 
previously in this context by Jaffe (1996) and, more recently, by
John \& Narlikar (2001).

\subsection{The Gaussian approximation to the posterior}
\label{gaussapprox}

Unfortunately, the evaluation of an evidence integral, such as 
(\ref{eviddef0}), is a challenging numerical task. From (\ref{pbardef}), we
see that 
\begin{equation}
\Pr(\vect{D}|H_0) = \int\overline{\Pr}(\btheta|\vect{D})\,d\btheta,
\label{evid2}
\end{equation}
and so the evidence may only be evaluated directly if one can calculate
$\overline{\Pr}(\btheta|\vect{D})$ over some hypercube in parameter
space, which we noted earlier is often computational unfeasible. 
Nevertheless, if the data are conclusive, we would expect this (unnormalised) 
posterior to be sharply peaked about the position of the maximum
$\btheta=\hat{\btheta}$. In this case, we may approximate this function
by performing a Taylor expansion about
$\btheta=\hat{\btheta}$. Working with the log-posterior, and
keeping only terms up to second-order in $\btheta$, 
leads to the Gaussian approximation of the (unnormalised) posterior
distribution, which reads
\begin{equation}
\overline{\Pr}(\btheta|\vect{D}) \approx 
\overline{\Pr}(\hat{\btheta}|\vect{D})\exp
\left[-\tfrac{1}{2}(\btheta-\hat{\btheta})^{\rm T}\vect{C}^{-1}
(\btheta-\hat{\btheta})\right],
\label{gaussapp}
\end{equation}
where $\vect{C}^{-1}$ is the estimated inverse 
covariance matrix of the parameters and is given by
\begin{equation}
\vect{C}^{-1} = -\left.\nabla\nabla\ln\Pr(\btheta|\vect{D})
\right|_{\btheta=\hat{\btheta}}
=-\left.\nabla\nabla\ln\overline{\Pr}(\btheta|\vect{D})
\right|_{\btheta=\hat{\btheta}}.
\end{equation}
Substituting the form (\ref{gaussapp}) into (\ref{evid2}), we thus find that an
approximation to the value of the evidence is given by
\begin{equation}
\Pr(\vect{D}|H_0) \approx (2\pi)^{M/2} |\vect{C}|^{1/2} \,
\Pr(\hat{\btheta})\Pr(\vect{D}|\hat{\btheta})
\label{evidapprox}
\end{equation}
where $M$ is the number of parameters of interest $\btheta$
and we have rewritten $\overline{\Pr}(\hat{\btheta}|\vect{D})$ using 
(\ref{pbardef}). Since
the estimation of parameter values and their associated errors already
requires one to calculate all the quantities on the right-hand side of
(\ref{evidapprox}), we see that 
this approximate evidence may be evaluated with no extra
work. We note, however, that for (\ref{evidapprox}) to hold, the prior
and likelihood must be {\em correctly normalised}, such that
$\int\Pr(\btheta)\,d\btheta=1$ and
$\int\Pr(\vect{D}|\btheta)\,d\vect{D}=1$. We also note, in this case, that
consideration of the ratio of evidences is a natural generalisation
of the standard likelihood-ratio approach to model comparison.

\subsection{Markov-Chain Monte-Carlo methods}

We note, in passing, that the approach to parameter estimation and
the approximate evaluation of evidences outlined above may soon
become obsolete. With the advent of faster computers and 
efficient algorithms, it has recently become numerically feasible to
sample directly from the posterior distribution using Monte-Carlo
Markov-Chain (MCMC) techniques (see Knox, Christensen \& Skordis
2001). This allows one trivially to
obtain one-dimensional marginalised posteriors for each parameter of
interest $a_i$, and also enables the direct numerical evaluation of
evidence integrals. Clearly, the MCMC technique has enormous
potential for the future estimation of cosmological parameters.

\section{Hyperparameters}
\label{hyper}

In this paper, we wish to construct a robust technique for performing
a joint estimation of cosmological parameters from combined datasets.
The basic idea behind the approach presented here is to introduce
additional {\em hyperparameters} $\balpha$ into the Bayesian inference
problem. In
other words, we extend our parameter vector to include not only the
parameters of interest $\btheta$, but also the hyperparameters
$\balpha$. These hyperparameters are analogous to `nuisance'
parameters that often arise in the standard approach to parameter
estimation. In our case, however, they are not present in our model
a priori, but we have chosen to introduce them in order to allow extra
freedom in the parameter estimation process. 

\subsection{Marginalisation over hyperparameters}

As with any set of
nuisance parameters, we must integrate out (or marginalise over) the
hyperparameters $\balpha$ in order to recover the posterior
distribution of our parameters of interest $\btheta$. 
Thus, we have
\begin{eqnarray}
\Pr(\btheta|\vect{D}) 
& = & \int\Pr(\btheta,\balpha|\vect{D})\,d\balpha  \nonumber \\
& = & \frac{1}{\Pr(\vect{D})}\int\Pr(\vect{D}|\btheta,\balpha)
\Pr(\btheta,\balpha)\,d\balpha. \label{eqn1}
\end{eqnarray}
In this paper, we shall assume that the parameters of interest
$\btheta$ and the hyperparameters $\balpha$ are independent, so
that 
\begin{equation}
\Pr(\btheta,\balpha) = \Pr(\btheta)\Pr(\balpha).
\label{aalpsep}
\end{equation}
Substituting
this form of the prior into (\ref{eqn1}), we immediately recover Bayes'
theorem (\ref{bayes}), where the form of the likelihood function in the
presence of hyperparameters is
\begin{equation}
\Pr(\vect{D}|\btheta) 
= \int\Pr(\vect{D}|\btheta,\balpha) \Pr(\balpha)\,d\balpha.
\label{hyperlike}
\end{equation}
Indeed, under the assumption (\ref{aalpsep}), 
this expression for the likelihood embodies the
{\em complete} hyperparameters technique.

Let us now turn our attention to the form of this likelihood function. 
Assuming that the totality of data $\vect{D}$ consists of $N$ {\em independent}
datasets $\vect{D}_k$ $(k=1,\ldots,N)$, we have
\begin{equation}
\Pr(\vect{D}|\btheta,\balpha)
=\prod_{k=1}^N \Pr(\vect{D}_k|\btheta,\balpha).
\label{eqn2}
\end{equation}
Moreover, in this paper, we will make the additional simplifying 
assumption that the $k$th 
term in the product on the right-hand side of (\ref{eqn2}) depends
only on $\alpha_k$. In this case, it reduces to
\begin{equation}
\Pr(\vect{D}|\btheta,\balpha)
=\prod_{k=1}^N \Pr(\vect{D}_k|\btheta,\alpha_k).
\end{equation}
We shall also assume that the individual hyperparameters $\alpha_i$ 
$(i=1,\ldots,N)$ are themselves independent, so that
$\Pr(\balpha) = \Pr(\alpha_1)\Pr(\alpha_2)\cdots\Pr(\alpha_N)$.  Thus
the expression (\ref{hyperlike}) for the likelihood becomes
\begin{equation}
\Pr(\vect{D}|\btheta) = \prod_{k=1}^N 
\int\Pr(\vect{D}_k|\btheta,\alpha_k) \Pr(\alpha_k)\,d\alpha_k,
\label{hyperlike2}
\end{equation}
which is simply the product of the individual likelihoods 
$P(\vect{D}_k|\btheta)$ for each dataset, 
after marginalisation over the hyperparameter.
This form can be substituted into Bayes' theorem (\ref{bayes}) 
to obtain the posterior $\Pr(\btheta|\vect{D})$, which can then be
used to obtain the best estimates $\hat{\btheta}$ of
the parameters and their associated errors.

Finally, it is worth noting that we may use (\ref{hyperlike2}) to write the
standard approach to parameter estimation, in which no hyperparameters 
are introduced, as a special case of the hyperparameters technique.
This is achieved by fixing the hyperparameters $\alpha_k$ 
to have particular values $\alpha_k^0$ $(k=1,\ldots,N)$, which 
is easily accommodated in the above formalism by assigning the priors
\begin{equation}
\Pr(\alpha_k)=\delta(\alpha_k-\alpha_k^0),
\end{equation}
where $\delta(z)$ is the Dirac delta function.
Most often, the hyperparameters are introduced in such a way that
$\alpha_k^0=1$ (for all $k$) corresponds to the standard approach, in
which hyperparameters are absent. In other words, the standard
likelihood function for the $k$th dataset may be denoted by
$\Pr(\vect{D}_k|\btheta,\alpha_k=1)$. 

\subsection{Hyperparameters and evidence}
\label{evidence}

So far, we have not addressed the question of whether the data support
the introduction of hyperparameters in the first instance. For example,
if a collection of different datasets are all
mutually-consistent, then one might not consider it wise to
introduce the hyperparameters $\balpha$. Indeed, their introduction 
could lead to larger uncertainties on the estimated values of the parameters of
interest $\btheta$, since one has to perform a marginalisation
over $\balpha$. On the other hand, if several of the datasets are
not in good agreement, the use of hyperparameters is necessary, in
order to obtain statistically meaningful results. 

In fact, Bayes' theorem itself 
allows us to make an objective decision
regarding whether the data warrant the introduction of
hyperparameters. As discussed in section~\ref{evidmodel}, this may be 
achieved using the Bayesian evidence
$\Pr(\vect{D})$. For illustration, let us
consider two models for the data as follows: 
the model $H_0$ does not include hyperparameters, whereas the
model $H_1$ assigns a free hyperparameter to each dataset.
In either case, the 
probability of obtaining the observed data $\vect{D}$ is
\begin{eqnarray}
\Pr(\vect{D})
& = &  \int \Pr(\vect{D},\btheta,\balpha)\,d\balpha\,d\btheta
\nonumber \\
& = &  \int \Pr(\vect{D}|\btheta,\balpha)\Pr(\btheta,\balpha)
\,d\balpha \,d\btheta \nonumber \\
& = &  \int \Pr(\vect{D}|\btheta)\Pr(\btheta) \,d\btheta,
\label{evidencedef}
\end{eqnarray}
where, in the last line, the likelihood $\Pr(\vect{D}|\btheta)$ is
given by (\ref{hyperlike2}). For $H_0$, the priors are simply 
$\Pr(\alpha_k)=\delta(\alpha_k-1)$ and
the integral simplifies accordingly, whereas for $H_1$ the priors
will, in general, have some more complicated form. 

Denoting the resulting evidence values for our two hypotheses by
$\Pr(\vect{D}|H_0)$ and $\Pr(\vect{D}|H_1)$ respectively,
the question of whether the data warrant the introduction of
hyperparameters, is now answered simply by comparing the relative
values of $\Pr(\vect{D}|H_0)$ and $\Pr(\vect{D}|H_1)$. 
The hypothesis
having the larger evidence is the one that should be accepted.
If $\Pr(\vect{D}|H_1) \la \Pr(\vect{D}|H_0)$, this
indicates that the datasets are all mutually consistent and
that the introduction of hyperparameters is not warranted. Conversely,
the condition $\Pr(\vect{D}|H_1) \gg \Pr(\vect{D}|H_0)$ is strong
indication that the datasets are {\em not} mutually consistent, and
that hyperparameters are necessary in order to obtain meaningful
statistical results.

\section{Weighting of datasets}
\label{weighting}

In the above discussion, we have specialised to the case where a
single hyperparameter is associated with each dataset. We have not
yet, however, made explicit how this hyperparameter enters 
the form of the modified likelihood $\Pr(\vect{D}|\btheta,\alpha_k)$ 
for each dataset. Neither have we fixed the form of the prior
$\Pr(\alpha_k)$ on each hyperparameter.
As  mentioned in the Introduction,
an obvious use of hyperparameters in cosmological parameter estimation
is in the weighting of the different datasets being analysed. We now
consider this application in more detail.

\subsection{The prior on each hyperparameter}
\label{weight_prior}

Let us first turn our attention to the prior $\Pr(\alpha_k)$ on each
hyperparameter. As discussed in Paper I, since $\alpha_k$ is a scale 
parameter, we might adopt 
the `non-informative' or Jeffrey's prior, which is uniform in the
logarithm of the parameter. Thus, we have
\begin{equation}
\overline{\Pr}(\alpha_k) = \frac{1}{\alpha_k},
\end{equation}
(or, more generally, one might take $\Pr(\alpha)=1/\alpha^n$, where
$n$ is a non-negative integer). 
The Jeffrey's prior does, however, cause some difficulties, since it is
improper and so cannot be normalised (as indicated).
This is not a problem for obtaining estimates
of the parameter values $\hat{\btheta}$ 
and their associated errors, since these
quantities do not depend on the overall normalisation of the posterior.
However, the use of an improper prior makes it impossible to
calculate evidences, and so we are unable to assess the whether the
data warrant the introduction of the hyperparameters.

We must ask ourselves, however, whether we are truly ignorant of the
value of each hyperparameter $\alpha_k$. In fact, this 
is rather unlikely. Given the
nature of the data-weighting problem, we might suppose that the
expectation value of $\alpha_k$ is unity. In other words, in the first
instance, we expect that the
experimental data have been correctly analysed and that the 
results are free from any systematic bias or misquoted random errors.
Thus, we need to assign a
prior probability distribution $\Pr(\alpha_k)$ subject to the
constraint that $E[\alpha_k]=1$. As shown by Jaynes (1957a,b), building
on the work of Shannon (1948), the only consistent way of assigning
the prior probability distribution $\Pr(\alpha_k)$ is by maximising
the `entropy' functional
\begin{equation}
S[\Pr(\alpha_k)] = -\int_0^\infty \Pr(\alpha_k)\ln\Pr(\alpha_k)\,d\alpha_k,
\label{entropy}
\end{equation}
subject to the normalisation 
constraint $\int_0^\infty \Pr(\alpha_k)\,d\alpha_k  = 1$ and 
the constraint $E[\alpha_k]=
\int_0^\infty \alpha_k \Pr(\alpha_k) \,d\alpha_k =  1$ on the expectation
value. This is a straightforward problem in the calculus of variations,
and has the simple solution
\begin{equation}
\Pr(\alpha_k) = \exp(-\alpha_k),
\label{expprior}
\end{equation}
which is, of course, properly normalised. 
In this paper, we shall use 
this exponential form of the prior.

We note in passing, however, that in some cases one may not only have 
$E[\alpha_k]=1$, but also some a priori expectation on the
{\em variance} of $\alpha_k$, i.e. some limit on the {\em range} of weights
that should be assigned to each dataset. In this case, we need to assign a
prior probability distribution $\Pr(\alpha_k)$ by maximising
(\ref{entropy}) subject to the constraints 
$\int_0^\infty \Pr(\alpha_k)\,d\alpha_k  = 1$, 
$E[\alpha_k]=1$ and $V[\alpha_k]=\sigma^2$ (say). Unfortunately, 
this problem has no solution as stated. Nevertheless, a closed-form
solution is easily obtained if one does not
restrict $\alpha_k$ to be non-negative, but instead allows it to
take any value between $-\infty$ and $\infty$ (thereby modifying the
limits on the integral in (\ref{entropy}) and in the normalisation of
$\Pr(\alpha_k)$). In this case, the calculus of variations problem
is easily solved to obtain the {\em Gaussian} form
\begin{equation}
\Pr(\alpha_k) = \frac{1}{\sqrt{2\pi}\sigma}\exp
\left[-\frac{(\alpha_k-1)^2}{2\sigma^2}\right],
\end{equation}
which is again properly normalised.
Clearly, this form is not strictly applicable to weights, which are
required to be non-negative. Nevertheless, provided $\sigma \ll 1$, the
above Gaussian form will be a good approximation to the true prior.

\subsection{The likelihood function for each dataset}
\label{weight_like}

If the hyperparameters $\alpha_k$ are to act as weights on the
different datasets, we specify the log-likelihood function for each dataset 
to have the form
\begin{equation}
\ln\Pr(\vect{D}_k|\btheta,\alpha_k) = 
\alpha_k\ln \Pr(\vect{D}_k|\btheta,\alpha_k=1)-\ln Z_k(\btheta,\alpha_k),
\label{loglikedef}
\end{equation}
where $\Pr(\vect{D}_k|\btheta,\alpha_k=1)$ 
is the standard likelihood function (in
the absence of hyperparameters) and $Z_k$ is a normalisation factor
that ensures $\int \Pr(\vect{D}_k|\btheta,\alpha_k)\,d\vect{D}_k=1$.
The expression (\ref{loglikedef}) 
for the log-likelihood clearly corresponds to the
likelihood function itself having the form
\begin{equation}
\Pr(\vect{D}_k|\btheta,\alpha_k) =
\frac{
\left[\Pr(\vect{D}_k|\btheta,\alpha_k=1)\right]^{\alpha_k}}
{Z_k(\btheta,\alpha_k)},
\label{likedef}
\end{equation}
from which we see that the normalisation factor has the explicit form
\begin{equation}
Z_k(\btheta,\alpha_k) = \int
\left[\Pr(\vect{D}_k|\btheta,\alpha_k=1)\right]^{\alpha_k}\,d\vect{D}_k.
\label{zdef}
\end{equation}
In particular, we note that $Z_k(\btheta,1)=1$.
From (\ref{loglikedef}), we see that the standard approach to joint parameter
estimation corresponds to taking $\alpha_k=1$ for those datasets that
are included in the analysis, and $\alpha_k=0$ for those that are discarded.
Using the prior (\ref{expprior}), and the form of the
likelihood for each dataset given in (\ref{likedef}), we find that the
new likelihood function for the $k$ dataset is given by
\begin{equation}
\Pr(\vect{D}_k|\btheta)
= \int_0^\infty \frac{
\left[\Pr(\vect{D}_k|\btheta,\alpha_k=1)\right]^{\alpha_k}}
{Z_k(\btheta,\alpha_k)}
e^{-\alpha_k}\,d\alpha_k.
\label{partlikedef}
\end{equation}

\subsection{The full likelihood and posterior distribution}
\label{weight_post}

Since the datasets $\vect{D}_k$ $(k=1,\ldots,N)$ are assumed to be
independent, the full likelihood function is given by
\begin{equation}
\Pr(\vect{D}|\btheta)= \prod_{k=1}^N \Pr(\vect{D}_k|\btheta),
\label{fulllikedef}
\end{equation}
where $\Pr(\vect{D}_k|\btheta)$ is given by (\ref{partlikedef}).
The posterior distribution $\Pr(\btheta|\vect{D})$
of the parameters of interest is then given by Bayes'
theorem (\ref{bayes}).

For the purposes of illustration, let us suppose finally that the prior
$\Pr(\btheta)$ on the parameters of interest is {\em uniform}. In
order that we may subsequently calculate evidences, we must ensure
that this prior is properly normalised. Suppose the prior is zero
outside some (large) region ${\cal R}$ is the $M$-dimensional
parameter space. Thus, we may define the prior as
\begin{equation}
\Pr(\btheta) = \left\{
\begin{array}{cl}
1/V({\cal R}) & \mbox{if $\btheta$ lies in ${\cal R}$} \\
0              & \mbox{otherwise},
\end{array}
\right.
\label{unipdef}
\end{equation}
where $V({\cal R}) = \int_{\cal R} \Pr(\btheta)\,d\btheta$ is the
`volume' of the region ${\cal R}$.

The resulting posterior distribution $\Pr(\btheta|\vect{D})$
can then be calculated over some $M$-dimensional hypercube in
parameter space, or used in the standard way to obtain 
the estimates $\hat{\btheta}$
and their associated errors. 
We note that, since the hyperparameters $\alpha_k$ have been
marginalised over in (\ref{fulllikedef}), they 
do not have specific values that can be quoted
at the end of the analysis. Nevertheless, it can be useful to know
which values of $\alpha_k$ were most favoured, and hence which
datasets were given a large or small weight in the analysis. 
Thus, at each point in the space of parameters $\btheta$, we define
the `effective' weight $\alpha_k^{\rm eff}(\btheta)$ to be
that which maximises the corresponding individual likelihood 
$\Pr(\vect{D}_k|\btheta,\alpha_k)$ at that point. Of course,
the most relevant set of such quantities are those evaluated at
the point $\btheta=\hat{\btheta}$.

\section{Gaussian likelihood functions}
\label{gausslike}

Our analysis has thus far been presented in a general form, which may
be applied to a wide range of problems in which one wishes to weight
different datasets. It is quite common, however, for the likelihood
function to take the form of a multivariate Gaussian in the data. 
In this case, for the $k$th dataset, we have
\begin{equation}
\Pr(\vect{D}_k|\btheta,\alpha_k=1) 
= \frac{1}{(2\pi)^{n_k/2}|\vect{V}_k|^{1/2}}
\exp\left(-\tfrac{1}{2}\chi_k^2\right),
\label{h0likea}
\end{equation}
where 
\begin{equation}
\chi_k^2 = (\vect{D}_k-\bmu_k)^{\rm T}\vect{V}_k^{-1} (\vect{D}_k-\bmu_k).
\label{chi2def}
\end{equation}
In these expressions, $n_k$ is the number of items in 
the $k$th dataset, $\bmu_k$ is
the expectation value of the quantities $\vect{D}_k$ and
$\vect{V}_k$ is their covariance matrix. In general, both $\bmu_k$ and
$\vect{V}_k$ may depend on the parameters of interest $\btheta$.
(note that, in general, the likelihood function is {\em not}
a multivariate Gaussian in the parameter space $\btheta$, unless 
$\vect{V}_k$ does not depend on the parameters $\btheta$ and
$\bmu_k$ depends only linearly on them).

We note that, even when the form of the likelihood is not Gaussian,
one can often perform a coordinate transformation to a new set of data
variables $\vect{z}_k$, for which the likelihood is (approximately)
Gaussian (see, for example, Bond, Jaffe \& Knox 2000). Moreover, as
shown by Bridle et al. (2001) in the context of CMB observations,
if the data $\vect{D}_k$ originally
obey a Gaussian likelihood function of the form (\ref{h0likea}), then
one can perform analytic marginalisations over calibration and
`beam' uncertainties that yield a new likelihood function which
is {\em also} of Gaussian form, but with a modified `covariance' matrix
$\vect{V}'_k$.

In the standard approach to estimating the parameters $\btheta$, one
assumes a model $H_0$, in which no hyperparameters
are introduced. Thus the full likelihood function is given simply by
\begin{equation}
\Pr(\vect{D}|\btheta,H_0) 
= \prod_{k=1}^N \frac{1}{(2\pi)^{n_k/2}|\vect{V}_k|^{1/2}}
\exp\left(-\tfrac{1}{2}\chi_k^2\right),
\label{h0like}
\end{equation}
where, on the left-hand side, we have made explicit the conditioning
on $H_0$. One then substitutes this expression into 
(\ref{pbardef}) to obtain the corresponding 
(unnormalised) posterior $\overline{\Pr}(\btheta|\vect{D},H_0)$.

Alternatively, we may adopt the model $H_1$, in which a free
hyperparameter $\alpha_k$ is assigned as a weight to each dataset.
Defining the modified likelihood function for each dataset
$\Pr(\vect{D}_k|\btheta,\alpha_k)$ by (\ref{likedef}), and 
using (\ref{zdef}) 
to evaluate the normalistion factor $Z_k(\btheta,\alpha_k)$, one finds
\begin{equation}
\Pr(\vect{D}_k|\btheta,\alpha_k) 
= \frac{1}{(2\pi)^{n_k/2}|\vect{V}_k|^{1/2}} \alpha_k^{n_k/2}
\exp\left(-\tfrac{1}{2}\alpha_k\chi_k^2\right),
\label{likealpha}
\end{equation}
This result may then be substituted into the expression
(\ref{partlikedef}) to obtain the new likelihood function for the
$k$th dataset, which is given by
\begin{equation}
\Pr(\vect{D}_k|\btheta) 
= \frac{1}{(2\pi)^{n_k/2}|\vect{V}_k|^{1/2}} \int_0^\infty \!\!\alpha_k^{n_k/2}
e^{-\alpha_k(\tfrac{1}{2}\chi_k^2+1)}\,d\alpha_k.
\end{equation}
The integral over $\alpha_k$ can be performed easily, using the definition
of the Gamma function $\Gamma(n) = \int_0^\infty x^{n-1}e^{-x}\,dx$,
and one finds
\begin{equation}
\Pr(\vect{D}_k|\btheta) = 
\frac{2\Gamma\left(\frac{n_k}{2}+1\right)}
{\pi^{n_k/2}|\vect{V}_k|^{1/2}} (\chi_k^2+2)^{-\left(\tfrac{n_k}{2}+1\right)},
\label{h1likea}
\end{equation}
which is properly normalised. Thus, assuming the datasets to be
independent, the full likelihood (\ref{fulllikedef}) is given by
\begin{equation}
\Pr(\vect{D}|\btheta,H_1) = 
\prod_{k=1}^N \frac{2\Gamma\left(\frac{n_k}{2}+1\right)}
{\pi^{n_k/2}|\vect{V}_k|^{1/2}} (\chi_k^2+2)^{-\left(\tfrac{n_k}{2}+1\right)},
\label{h1like}
\end{equation}
where we have made explicit the conditioning on $H_1$.
As above, this form of the likelihood
can then be substituted into (\ref{pbardef}) to obtain the corresponding
(unnormalised) posterior $\overline{\Pr}(\btheta|\vect{D},H_1)$. 

\subsection{Evaluation of the posterior and evidences}

Ideally one would wish to calculate the full (unnormalised) posteriors
$\overline{\Pr}(\btheta|\vect{D},H_0)$ and 
$\overline{\Pr}(\btheta|\vect{D},H_1)$ on some hypercube in parameter space.
In this way, the location of the (global) maximum is obtained
immediately, and the presence of multiple peaks in the posterior(s) is
readily observed. Moreover, marginalised distributions may be
trivially calculated, in order to place confidence limits on
individual parameter values. Assuming the uniform prior
(\ref{unipdef}), we see from (\ref{h0like})
and (\ref{h1like}) that the evaluation of the posterior for the
models $H_0$ and $H_1$ respectively requires similar functions
to be evaluated. In other words, if one has an algorithm for
calculating $|\vect{V}_k|$ and $\chi_k^2$, as required for the
evaluation of the standard posterior (model $H_0$), then
one can immediately evaluate the hyperparameters posterior (model
$H_1$). 

An additional advantageous feature of calculating the full 
(unnormalised) posteriors $\overline{\Pr}(\btheta|\vect{D},H_0)$ and 
$\overline{\Pr}(\btheta|\vect{D},H_1)$ on a hypercube in parameter
space is that it allows the immediate evaluation of the evidence in
each case, which are given by
\begin{eqnarray}
\Pr(\vect{D}|H_0) & = & \int \overline{\Pr}(\btheta|\vect{D},H_0) 
\,d\btheta, \label{h0evid}\\
\Pr(\vect{D}|H_1) & = & \int \overline{\Pr}(\btheta|\vect{D},H_1)
\,d\btheta. \label{h1evid}
\end{eqnarray}
The ratio of these quantities can then be used to decide whether or
not the inclusion of hyperparameters is warranted by the data.

\subsection{Estimation of parameter values}
\label{gauss_est}

Unfortunately, evaluation of the full posterior
distribution(s) on a hypercube in parameter space is only
feasible when the number of parameters $M$ is small (although, as
mentioned in section~\ref{gaussapprox}, 
this restriction may be overcome using MCMC sampling techniques).
For larger problems, the estimates $\hat{\btheta}$ of the
parameters of interest are usually obtained by maximising the posterior. 
Assuming the uniform prior (\ref{unipdef}), for the standard model
$H_0$ (without hyperparameters)  this corresponds to
minimising the function
\begin{equation}
-2\ln\Pr(\btheta|\vect{D},H_0) = 
\sum_{k=1}^N \left(\ln|\vect{V}_k|+ \chi_k^2\right)
+ c,
\label{h0loglike}
\end{equation}
where $c$ is a constant. This is, of course, a very familiar
result. For
the model $H_1$, however, which contains hyperparameters, one must
instead minimise 
\begin{equation}
-2\ln\Pr(\btheta|\vect{D},H_1) = 
\sum_{k=1}^N \left[\ln|\vect{V}_k|+ 
(n_k+2) \ln(\chi_k^2+2)\right]
+ c,
\label{h1loglike}
\end{equation}
where $c$ is a (different) constant.

From (\ref{h0loglike}), for the model
$H_0$, the parameters estimates $\hat{\btheta}_0$ may be obtained by solving
\begin{equation}
\sum_{k=1}^N (\nabla\ln|\vect{V}_k| + \nabla\chi_k^2) =
0,
\label{a0fromh0}
\end{equation}
whereas for the model $H_1$, the estimates $\hat{\btheta}_1$
are found by solving
\begin{equation}
\sum_{k=1}^N [\nabla\ln|\vect{V}_k| + 
(n_k+2) \nabla\ln(\chi_k^2+2)] =
0.
\label{a1fromh1}
\end{equation}
In the last case, however, we may write
\begin{equation}
\nabla\ln(\chi_k^2+2) = (\nabla\chi_k^2)/(\chi_k^2+2).
\end{equation}
Thus, if one is able to evaluate the derivatives
$\nabla\ln|\vect{V}_k|$ and $\nabla\chi_k^2$ for $k=1,\ldots,N$, 
to obtain the standard parameter estimates $\hat{\btheta}_0$, one
may easily obtain the estimates $\hat{\btheta}_1$.

We note, in passing, that in the special case where the covariance matrices 
$\vect{V}_k$ do not depend on the parameters $\btheta$, the 
expression (\ref{h1loglike}) 
is very similar to that presented in Paper I. In that paper,
an improper Jeffrey's prior was assumed on the hyperparameters, and
it was found that the parameter estimates $\hat{\btheta}_1$ were given by
minimising the quantity
\begin{equation}
\sum_{k=1}^N n_k \ln (\chi_k^2).
\label{old}
\end{equation}
Unfortunately, the 
corresponding posterior in this case cannot be normalised, as a
result of using an improper prior. Nevertheless, in the limit 
where the number of data items $n_k$ in each
dataset is large, we would expect $\chi_k^2$ also to be large. Thus,
in this limit, minimisation of the 
expressions (\ref{h1loglike}) and (\ref{old}) 
respectively (in the case where the
$\vect{V}_k$ are independent of the parameters $\btheta$)
should give almost identical parameter estimates $\hat{\btheta}_1$.

As mentioned in section~\ref{weight_post}, since we have marginalised over the
hyperparameters $\alpha_k$ $(k=1,\ldots,N)$ in (\ref{h1loglike}), 
we cannot quote a value
for them at the end of the analysis, but we can determine an
`effective' value $\alpha_k^{\rm eff}(\btheta)$ for each
hyperparameter at any point in parameter space.
In
each case, this is given by the value of $\alpha_k$ that maximises the
modified likelihood $\Pr(\vect{D}|\btheta,\alpha_k)$ at that point.
Differentiating the 
expression (\ref{likealpha}) with respect to $\alpha_k$
and setting the result equal to zero, we find
\begin{equation}
\alpha_k^{\rm eff}(\btheta) = \frac{n_k}{\chi_k^2(\btheta)}.
\label{aeffdef}
\end{equation}
Clearly, this expression is most meaningful when 
evaluated at the point $\btheta=\hat{\btheta}$.

Finally, it is worth noting the case in which minimisation of the functions
(\ref{h0loglike}) and (\ref{h1loglike}) gives the {\em same} parameters
estimates $\hat{\btheta}$. We see that the 
expressions (\ref{a0fromh0}) and (\ref{a1fromh1}) 
are identical when
$\chi_k^2(\hat{\btheta})=n_k$ for all $k$.
From (\ref{aeffdef}), we see that this
corresponds to the case in which $\alpha_k^{\rm eff}(\hat{\btheta})=1$ for all
$k$.

\subsection{Error estimates and approximate evidences}
\label{gauss_evid}

The estimated covariance matrix of the parameter errors is given by
(\ref{covmatdef}). From (\ref{h0loglike}) and (\ref{h1loglike}), 
for the models $H_0$ and $H_1$ respectively, we have
\begin{eqnarray}
\vect{C}_0^{-1} & = & 
\tfrac{1}{2}\sum_{k=1}^N
[\nabla\nabla\ln|\vect{V}_k|+\nabla\nabla\chi_k^2]_{\btheta
=\hat{\btheta}_0}, \\
\vect{C}_1^{-1} & = & 
\tfrac{1}{2}\sum_{k=1}^N
[\nabla\nabla\ln|\vect{V}_k|+(n_k+2) \nabla\nabla
\ln(\chi_k^2+2)]_{\btheta=\hat{\btheta}_1}.
\end{eqnarray}
Since we may write
\begin{equation}
\nabla\nabla
\ln(\chi_k^2+2) = \frac{(\chi_k^2+2)\nabla\nabla\chi_k^2-
(\nabla\chi_k^2)(\nabla\chi_k^2)^{\rm T}}
{(\chi_k^2+2)^2},
\end{equation}
we see that, if one can calculate
the functions $\nabla\nabla\ln|\vect{V}_k|$
and $\nabla\nabla\chi_k^2$ (together with $\nabla\chi_k^2$, which  
is required for obtaining the parameter estimates) to
give the standard (inverse) covariance matrix $\vect{C}_0^{-1}$, one
may easily calculate $\vect{C}_1^{-1}$.

Once the (inverse) covariance matrices of the errors have been calculated,
the evaluation of (approximate) evidences is straightforward. Using
the approximate expression (\ref{evidapprox}) for the evidence, and
assuming the uniform prior (\ref{unipdef}), we see 
that the ratio of evidences for the two
models $H_0$ and $H_1$ is given by
\begin{equation}
\frac{\Pr(\vect{D}|H_1)}{\Pr(\vect{D}|H_0)}
\approx 
\frac{|\vect{C}_1|^{1/2}\, \Pr(\vect{D}|\hat{\btheta}_1,H_1)}
{|\vect{C}_0|^{1/2} \,\Pr(\vect{D}|\hat{\btheta}_0,H_0)}.
\label{approxevrat}
\end{equation}
If this ratio is much less than unity, one concludes that the
datasets are mutually consistent and do not support the 
introduction of hyperparameters; one should then quote
the results of the standard analysis, given by the
estimates $\hat{\btheta}_0$ and the covariance matrix $\vect{C}_0$.
If the ratio is much greater than unity, however, then this indicates
that inconsistencies do exist between (some of) the datasets and that
the inclusion of
hyperparameters is warranted; it is then more appropriate to quote
the estimates $\hat{\btheta}_1$ and the covariance matrix $\vect{C}_1$.
\begin{figure*}
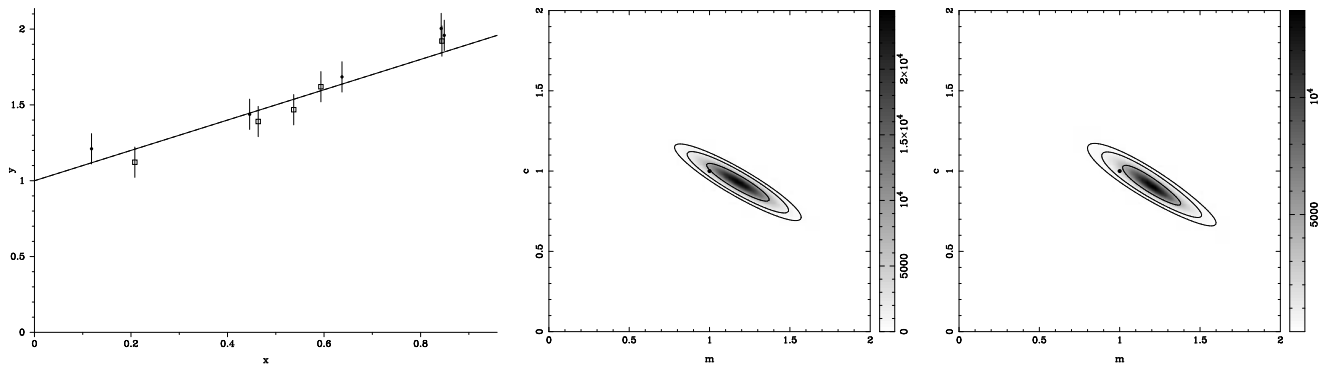

\begin{center}
\centerline{
\epsfig{file=fig1a.ps,angle=-90,width=6.5cm}
\quad
\epsfig{file=fig1b.ps,angle=-90,width=5.1cm}
\quad
\epsfig{file=fig1c.ps,angle=-90,width=5.1cm}
}
\caption{Left: the two data sets $\vect{D}_1$ (solid circles) 
and $\vect{D}_2$ (open squares) drawn from the straight line model
with slope $m=1$ and intercept $c=1$ (solid line), and subject to independent 
Gaussian noise of rms $\sigma_1=0.1$ and $\sigma_2=0.1$ respectively.
Middle: the (unnormalised) posterior
$\overline{\Pr}(\btheta|\vect{D},H_0)$ corresponding to the standard
approach to parameter estimation. Right:  the (unnormalised) posterior
$\overline{\Pr}(\btheta|\vect{D},H_1)$ corresponding to the
hyperparameters approach. In each case, the true parameter values are
indicated by the solid circle, and  the contours contain 68, 95 and
99 per cent respectively of the total probability.}
\label{fig1}
\end{center}
\end{figure*}

\section{A toy problem}
\label{toy}

To illustrate the potential usefulness of the hyperparameters approach
to weighting datasets, in this section we apply the method to the classic
parameter estimation problem of fitting a straight line through a set
of data points. In particular, we will show how the use of
hyperparameters as weights can overcome the common problems of
inaccurately quoted error-bars and the presence of systematic errors in
the measurements (see also Bridle 2000).

Let us suppose that the true underlying model for some process is the
straight line
\begin{equation}
y(x)=mx+c,
\label{sline}
\end{equation}
where the slope $m=1$ and the intercept $c=1$. We assume that two
independent sets of measurements $\vect{D}_1$ and $\vect{D}_2$
are made, each of which contains five data
points, so $n_1=5$ and $n_2=5$. In simulating each dataset, 
the $x$-values are drawn at random from a uniform distribution between
zero and unity, and we assume that these values are known precisely.
The corresponding $y$-values are 
independently distributed about their true model values according to 
a Gaussian distribution of known variance $\sigma_k$
for each data set. Thus, the likelihood function for the $k$th dataset has
the Gaussian form given by (\ref{h0likea}) and (\ref{chi2def}), where
$(\bmu_{k})_j = m(\vect{x}_k)_j + c$ and 
$\vect{V}_k=\mbox{diag}(\sigma_k^2,\ldots,\sigma_k^2)$. Thus, in this
simple example, the covariance matrix does not depend on the
parameters $m$ and $c$. 

We now consider three different scenarios,
which illustrate the general useful properties the hyperparameters
technique for weighting datasets. We note that, for this simple toy
problem, it is clear from a plot of the data points if inconsistencies
exist between two datasets, and we will see that the hyperparameters
approach verifies our initial suspicions where appropriate. Indeed,
this toy model was chosen so that inconsistencies revealed by the
hyperparameters method are transparent from a plot of the data.
It should be remembered, however, that for realistic cosmological
datasets and theoretical models, the situation is usually much more
complicated. In this case, one is often unable visually to discern any
inconsistencies in the data, but these may nevertheless be deduced by
using the hyperparameters technique. This is illustrated in
section~\ref{cosmoapp}.

\begin{figure*}
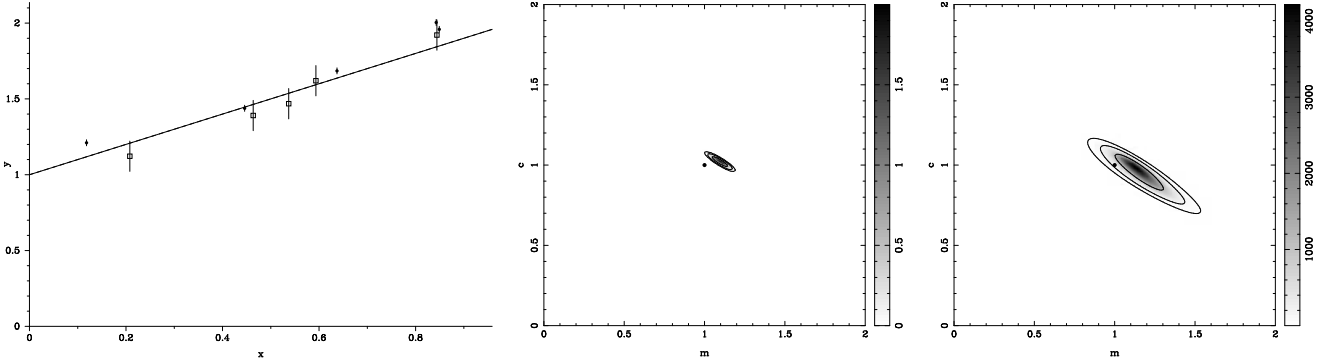

\begin{center}
\centerline{
\epsfig{file=fig2a.ps,angle=-90,width=6.5cm}
\quad
\epsfig{file=fig2b.ps,angle=-90,width=5.1cm}
\quad
\epsfig{file=fig2c.ps,angle=-90,width=5.1cm}
}
\caption{As for Fig.~\ref{fig1}, but the reported error-bars on the
dataset $\vect{D}_1$ have been underestimated by a factor of 5.}
\label{fig2}
\end{center}
\end{figure*}

\subsection{Accurate error-bars and no systematic error}

In our first case, both data sets $\vect{D}_1$ and
$\vect{D}_2$ are drawn from the correct model $m=1$ and $c=1$. The
noise rms for each dataset is $\sigma_1=\sigma_2=0.1$, and 
these (correct) values are used in the parameter estimation process.
The two simulated datasets are shown in Fig.~\ref{fig1} (left panel),
in which the dataset $\vect{D}_1$ is denoted by the filled circles,
whereas $\vect{D}_2$ corresponds to the open squares. The true
underlying model is plotted as a solid line. 

Assuming an appropriately normalised uniform prior on the parameters
in the range $m: 0 \to 2$ and $c: 0\to 2$, and using the 
expressions (\ref{h0like}) and (\ref{h1like}) for the 
likelihood in each case, we may calculate the (unnormalised) posteriors 
$\overline{\Pr}(m,c|\vect{D},H_0)$ and $\overline{\Pr}(m,c|\vect{D},H_1)$
corresponding to the standard and hyperparameters approaches
respectively (where $\vect{D}$ denotes the combination of the
datasets $\vect{D}_1$ and $\vect{D}_2$). These posterior distributions
are plotted in Fig.~\ref{fig1} (middle and right panels), in which the
contours contain 68, 95 and 99 per cent respectively of the 
total probability; the true parameter values are indicated by a solid
circle. 

In both cases $H_0$ and $H_1$, the posterior distributions
appear very similar and contain the true parameter values at about 
the 68 per cent confidence level. However, since the prior
and the likelihood function in each case are properly normalised, the
greyscale units on the plots in Fig.~\ref{fig1} are {\em not} 
arbitrary. Indeed, the
evidence in each case can be calculated directly by integrating numerically
under these distributions, and we find
\[
\frac{\Pr(\vect{D}|H_1)}{\Pr(\vect{D}|H_0)} = 0.54.
\]
This indicates that standard approach is very slightly
preferred or, equivalently, that the 
introduction of hyperparameters is marginally
disfavoured by the data.
An evidence ratio of order unity does not, however, provide
a strong indication that one case is preferred over the other. We may
conclude that, in this case, 
our ability to infer the parameter values $m$ and $c$ from the data
is essentially unaffected by the introduction of hyperparameters.

It is interesting to compare the exact evidence ratio above with that
obtained using the approximate expression (\ref{approxevrat}), in which each
posterior distribution is approximated by a Gaussian centred at its
peak. Since the function (\ref{sline}) is linear in the parameters
$m$ and $c$, and the covariance matrices $\vect{V}_k$ do not depend on
these parameter, the posterior $\overline{\Pr}(\btheta|\vect{D},H_0)$
is, in fact, truly Gaussian. We found, in this case, that the approximate
expression (\ref{evidapprox}) for the evidence did indeed agree with that
obtained by direct numerical integration. In the
hyperparameters case, however, the posterior
$\overline{\Pr}(\btheta|\vect{D},H_1)$ is {\em not} Gaussian. Nevertheless, we
found that the approximate expression (\ref{evidapprox}) 
underestimated the true
evidence value by only 2 per cent, and the approximate evidence ratio was
found to be 0.53.

Finally, we note the effective weight
assigned to each dataset at the peak $\hat{\btheta}_1$ of the
hyperparameters posterior $\overline{\Pr}(\btheta|\vect{D},H_1)$.
These values are given by (\ref{aeffdef}) evaluated at this point, and we  
find $\alpha_1^{\rm eff}(\hat{\btheta}_1)=1.79$ and
$\alpha_2^{\rm eff}(\hat{\btheta}_1)=2.64$.

\subsection{Inaccurate error-bars and no systematic error}

In this case, the datasets $\vect{D}_1$ and $\vect{D}_2$ are identical
to those used in the previous subsection, but we assume that the quoted
errors-bars on the dataset $\vect{D}_1$ have been underestimated by a
factor of 5, while the error-bars on the dataset $\vect{D}_2$ are
quoted correctly.  Thus, in the parameter estimation procedure, we assume
the (incorrect) values $\sigma_1=0.02$ and $\sigma_2=0.1$. The
data points and their reported error-bars are shown in Fig.~\ref{fig2}
(left panel).

The resulting (unnormalised) posteriors 
$\overline{\Pr}(\btheta|\vect{D},H_0)$ and 
$\overline{\Pr}(\btheta|\vect{D},H_1)$ are shown in the middle and
right-hand panels of the figure. In this case, the two
posteriors are very different. In the standard approach $H_0$, the
posterior distribution is tightly constrained about its
maximum as a result of underestimating the errors on dataset
$\vect{D}_1$. Indeed, this posterior is virtually indistinguishable
from that calculated from the dataset $\vect{D}_1$ alone. The
true parameter values are excluded at a confidence level that far
exceeds the 99 per cent limit. In 
the hyperparameters case, however, the posterior
is much broader and resembles the corresponding
distribution in Fig.~\ref{fig1}.  In particular, the true
parameter values are comfortably contained 
within the 95 per cent confidence limit.

Integrating under the posterior distributions directly, the
exact evidence ratio is found to be
\[
\frac{\Pr(\vect{D}|H_1)}{\Pr(\vect{D}|H_0)} = 2.1 \times 10^{4},
\]
which clearly implies that the data favour the introduction of
hyperparameter weights. Using the expression (\ref{approxevrat}), 
the approximate evidence ratio is $1.6 \times 10^{4}$, which shows that
the Gaussian approximation to the hyperparameters posterior is once again
reasonably accurate. 

At the peak $\hat{\btheta}_1$
of the hyperparameters posterior, we find the effective weights
assigned to the two datasets $\vect{D}_1$ and $\vect{D}_2$ to be
$\alpha_1^{\rm eff}(\hat{\btheta}_1)=0.12$ and
$\alpha_2^{\rm eff}(\hat{\btheta}_1)=1.28$. Thus, the
first dataset (with error-bars underestimated by a factor of 5) has
been assigned an appropriate smaller statistical weight.

\subsection{Accurate error-bars and a systematic error}

\begin{figure*}
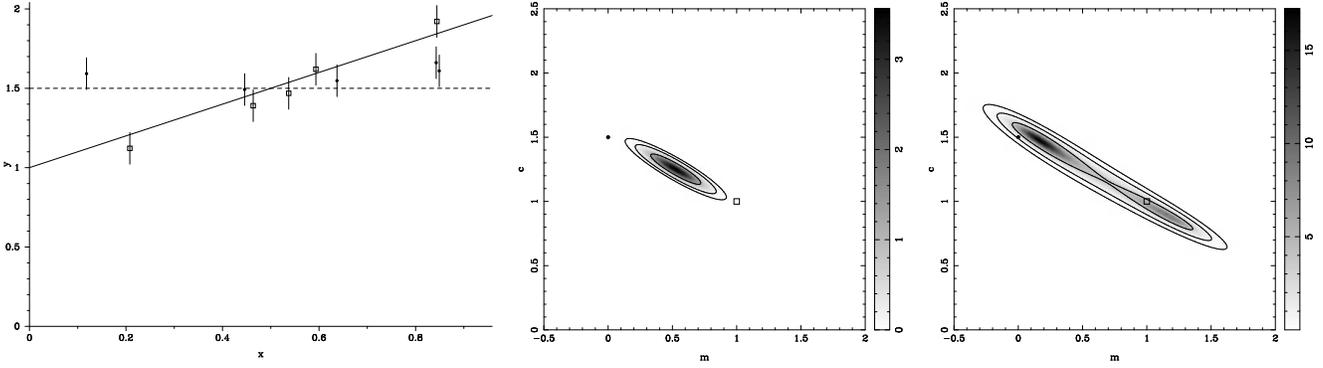

\begin{center}
\centerline{
\epsfig{file=fig3a.ps,angle=-90,width=6.5cm}
\quad
\epsfig{file=fig3b.ps,angle=-90,width=5.1cm}
\quad
\epsfig{file=fig3c.ps,angle=-90,width=5.1cm}
}
\caption{Left: the dataset $\vect{D}_1$ (solid circles) 
drawn from the straight-line model
with slope $m=0$ and intercept $c=1.5$ (dashed line) and the
dataset $\vect{D}_2$ (open squares) drawn from the straight-line model
with slope $m=1$ and intercept $c=1$ (solid line); each dataset
is subject to independent 
Gaussian noise of rms $\sigma_1=0.1$ and $\sigma_2=0.1$ respectively.
Middle: the (unnormalised) posterior
$\overline{\Pr}(\btheta|\vect{D},H_0)$ corresponding to the standard
approach to parameter estimation. Right:  the (unnormalised) posterior
$\overline{\Pr}(\btheta|\vect{D},H_1)$ corresponding to the
hyperparameters approach. In each case, the true parameter values 
for dataset $\vect{D}_1$ are 
indicated by the solid circle and the true parameter values for
dataset $\vect{D}_2$ are indicated by an open square.
The contours in each case contain 68, 95 and
99 per cent respectively of the total probability.}
\label{fig3}
\end{center}
\end{figure*}
\begin{figure*}
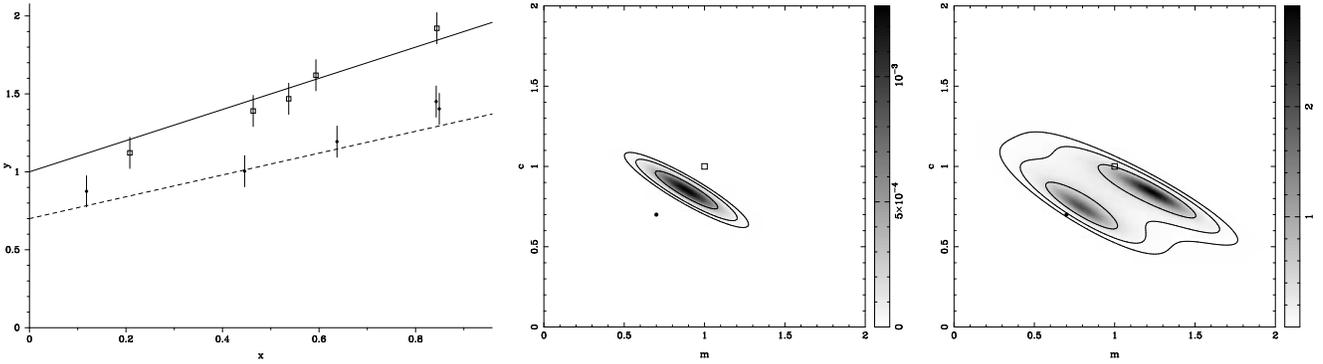

\begin{center}
\centerline{
\epsfig{file=fig4a.ps,angle=-90,width=6.5cm}
\quad
\epsfig{file=fig4b.ps,angle=-90,width=5.1cm}
\quad
\epsfig{file=fig4c.ps,angle=-90,width=5.1cm}
}
\caption{As for Fig.~\ref{fig3}, except that the dataset $\vect{D}_1$
(solid circles) 
is drawn from the straight-line model with slope $m=0.7$ and intercept
$c=0.7$.}
\label{fig4}
\end{center}
\end{figure*}

We now consider the introduction of a systematic error
into the dataset $\vect{D}_1$. This is simulated by drawing this
dataset from an incorrect straight-line model, for which the parameter values 
$m$ and $c$ differ from unity. Dataset $\vect{D}_2$, however, is still drawn
from the correct straight-line model with $m=1$ and $c=1$.
We shall, in fact, consider two separate
complementary cases. In the first case, we introduce a systematic
error in the direction of the natural degeneracy line in the $(m,c)$-plane,
whereas, in the second case, the introduced systematic error is
orthogonal to the natural degeneracy line. In both cases, we
assume that the error-bars on each dataset are quoted accurately as
$\sigma_1=0.1$ and $\sigma_2=0.1$ respectively.

\subsubsection{Case 1} 

In our first case, dataset $\vect{D}_1$ is drawn from a straight line
with slope $m=0$ and intercept $c=1.5$. The datasets
$\vect{D}_1$ and $\vect{D}_2$ are shown in Fig.~\ref{fig3} (left
panel), together with the underlying straight-line models from which each is
drawn.
The resulting posterior distributions
$\overline{\Pr}(\btheta|\vect{D},H_0)$ and 
$\overline{\Pr}(\btheta|\vect{D},H_1)$ are shown in the middle and
right-hand panels of the figure. Once again, 
the two
posteriors are very different. In the standard approach, $H_0$, the
posterior distribution peaks between the two sets of true values.
In spite of the fact that the two sets of parameter values define
a direction along
the natural degeneracy line in the $(m,c)$-plane, neither is contained
within the 99 per cent confidence contour, and so both models are 
excluded at a high significance level. 
In the hyperparameters case, however, the posterior
is much further extended along the natural degeneracy line. In particular, we
note that the distribution is bimodal, with each peak lying close to
one of the true sets of parameter values. Thus, the hyperparameters
indicates the presence of two underlying models for the
data, which signals an inconsistency between the two datasets.
This could be interpreted as one (or both) of the datasets containing a
systematic error.

The exact evidence ratio in this case is found to be
\[
\frac{\Pr(\vect{D}|H_1)}{\Pr(\vect{D}|H_0)} = 11.6,
\]
which gives a reasonably robust indication that
the data favour the introduction of
hyperparameters. Using the expression (\ref{approxevrat}), 
the approximate evidence ratio is given by 4.8. The reason for
the large inaccuracy in this case is that the Gaussian approximation to
the bimodal hyperparameters posterior is clearly rather
poor. In fact, as might be expected in this case, the Gaussian approximation
underestimates the true value of the evidence by about a factor of two.

Although the hyperparameters posterior is bimodal, the global 
maximum $\hat{\btheta}_1$ 
occurs at the peak close to the parameter values from which the
dataset $\vect{D}_1$ was drawn. At this peak, we find the effective weights
assigned to the two datasets $\vect{D}_1$ and $\vect{D}_2$ are
$\alpha_1^{\rm eff}(\hat{\btheta}_1)=3.11$ and
$\alpha_2^{\rm eff}(\hat{\btheta}_1)=0.18$, which indicates
(correctly) that the first dataset has been assigned a larger 
statistical weight at this point in parameter space.  However, at the
subsidiary peak $\hat{\btheta}'_1$ located near the parameter values
from which the dataset $\vect{D}_2$ was drawn, we find
$\alpha_1^{\rm eff}(\hat{\btheta}'_1)=0.11$ and
$\alpha_2^{\rm eff}(\hat{\btheta}_1)=6.41$, and so the roles of the
datasets have been reversed.

\subsubsection{Case 2}

In our second case, dataset $\vect{D}_1$ is drawn from a straight line
with slope $m=0.7$ and intercept $c=0.7$. The datasets
$\vect{D}_1$ and $\vect{D}_2$ are shown in Fig.~\ref{fig4} (left
panel), together with the underlying straight-line models from which each is
drawn.
The the middle and right-hand panels of the figure 
show the resulting posterior distributions
$\overline{\Pr}(\btheta|\vect{D},H_0)$ and 
$\overline{\Pr}(\btheta|\vect{D},H_1)$.
As in Case 1, the standard approach produces
a posterior distribution that peaks between the two true sets of
parameters values, excluding both at a high significance level. 
We note, in this case, that the two sets of true parameter
values define a direction orthogonal to the natural degeneracy
line in the $(m,c)$-plane.
In the hyperparameters case, however, the posterior is again bimodal, 
with each peak lying close to
one of the true sets of parameter values. Thus, once again the
hyperparameters approach signals the presence of two underlying models
for the data and hence an inconsistency between the datasets.

The exact evidence ratio is found to be
\[
\frac{\Pr(\vect{D}|H_1)}{\Pr(\vect{D}|H_0)} = 5.9 \times 10^{3},
\]
which strongly implies that the data favour the introduction of
hyperparameters. In this case, the approximate evidence ratio 
(\ref{approxevrat}) is given by $2.4 \times 10^{3}$. As in Case 1, 
the Gaussian approximation underestimates 
the true value of the evidence for the hyperparameters
posterior by about a factor of 2, as a result of it being bimodal.

Once again the effective weights clearly
show which dataset is dominating at each peak of the hyperparameters 
posterior.
At the global maximum $\hat{\btheta}_1$,
we find $\alpha_1^{\rm eff}(\hat{\btheta}_1)=0.066$ and
$\alpha_2^{\rm eff}(\hat{\btheta}_1)=7.96$, whereas at the subsidiary
maximum $\hat{\btheta}'_1$ we obtain
$\alpha_1^{\rm eff}(\hat{\btheta}'_1)=3.31$ and
$\alpha_2^{\rm eff}(\hat{\btheta}_1)=0.079$.

\section{A cosmological illustration}
\label{cosmoapp}

\begin{figure}
\begin{center}
\centerline{
\epsfig{file=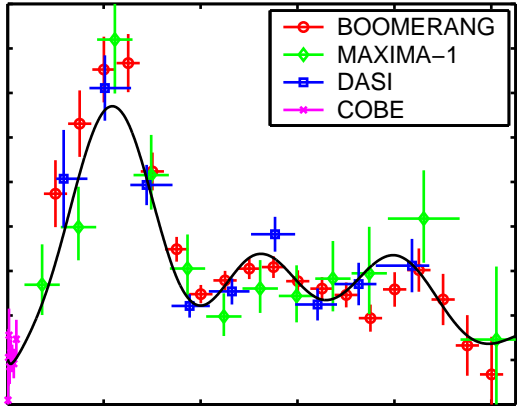,width=7cm}
}
\caption{The flat-band power estimates of the CMB power spectrum
reported by the COBE, BOOMERANG, MAXIMA, DASI experiments.
The solid line corresponds to the CMB power
spectrum for a spatially-flat inflationary CDM model with no 
tensor contribution and $\Omega_m=0.3$, $\Omega_b h^2=0.02$, 
$h=0.7$, $n=1$, $\tau=0$ and $Q=18$ $\mu$K.}
\label{cmbdata}
\end{center}
\end{figure}

In this section, we illustrate the hyperparameters technique 
for weighting datasets by applying it to 
recent measurements of the CMB power
spectrum provided by the BOOMERANG (Netterfield et al. 2001), 
MAXIMA (Lee et al. 2001) and DASI (Halverson et al. 2001) 
experiments. We also include the 8 data points for 
large-scale normalisation of the spectrum
provided by the COBE satellite. The flat band-powers reported by
each of these experiments are plotted in Fig~\ref{cmbdata}.
The solid line in the plot corresponds to the predicted CMB power
spectrum for a spatially-flat inflationary CDM model with no 
tensor contribution and $\Omega_m=0.3$, $\Omega_b h^2=0.02$, 
$h=0.7$, $n=1$, $\tau=0$ and $Q=18$ $\mu$K.

From Fig.~\ref{cmbdata}, we see immediately that there is 
good agreement between the datasets, all of which are broadly consistent
with the plotted theoretical CMB power spectrum. One would therefore
expect the inclusion of and marginalisation over hyperparameter
weights not to be warranted by the data. It should also be
remembered, however, that the points and error-bars plotted in the
figure do {\em not} include the calibration and beam uncertainties
associated with each experiment. Indeed, since these uncertainties can be as
large as 20 per cent in some cases, one would naively expect the
envelope of the experimental data to be much wider, resulting in 
poorer agreement with the theoretical spectrum. As we will see below,
this naive observation is supported by the proper calculation
of Bayesian evidences.

\subsection{No calibration or beam uncertainty}

We first analyse the data {\em without} including calibration and
beam errors, i.e. as plotted in Fig.~\ref{cmbdata}.
For each dataset $\vect{D}_k$ $(k=1,2,3,4)$, we assume the 
uncertainties on the flat band-powers are described by a multivariate
Gaussian of the form given in (\ref{h0likea}). For the COBE and DASI
datasets, we use the publicly-available window functions and
full covariance matrix $\vect{V}_k$ 
for each dataset. In the absence of the equivalent
information for the BOOMERANG and MAXIMA experiments, 
we assume a top-hat window 
function for the spectrum in each bin and neglect correlations between
bins. This approach is a good approximation of the correct one (see de
Bernardis et al. 2001) and does not affect our conclusions. 

Denoting the totality of the resulting data by the vector $\vect{D}$, the next
step in the analysis is to calculate the unnormalised posterior
distributions $\overline{\Pr}(\btheta|\vect{D},H_0)$ and 
$\overline{\Pr}(\btheta|\vect{D},H_1)$,
corresponding to the standard and hyperparameters
approaches respectively, for some set of cosmological parameters
$\btheta$. For the purposes of illustration, 
we assume a spatially-flat Universe ($\Omega_k=1$), with no
tensor contribution to the CMB spectrum, and take the parameter 
vector $\btheta$ to consist of 5 parameters, namely the
normalisation $Q$, the Hubble parameter $h$, the matter density 
$\Omega_{\rm m}$, the physical baryon density $\Omega_{\rm b}h^2$ and
the scalar spectral index $n$. Since the parameter space is only
5-dimensional, it is feasible to evaluate the 
unnormalised posteriors over a hypercube. 
Thus, we assume (suitably normalised) uniform priors on the parameters,
with the ranges $12  < Q < 24$ $\mu$K, $0.5 < h < 0.9$, 
$0.1 < \Omega_{\rm m} < 0.5$, $0.01 < \Omega_{\rm b}h^2 < 0.04$ and
$0.8 < n < 1.2$. The corresponding likelihood functions for the
hypotheses $H_0$ and $H_1$ have the forms (\ref{h0like}) and
(\ref{h1like}) respectively. The posterior distribution 
is calculated at 20 points in 
the directions $Q$, $h$ and $\Omega_m$ in parameter space, and
at 10 points in the directions $n$ and $\Omega_bh^2$.
The theoretical
power spectrum corresponding to each model
was calculated using CAMB (Lewis, Challinor \& Lasenby 2000).
\begin{figure*}
\begin{center}
\centerline{
\epsfig{file=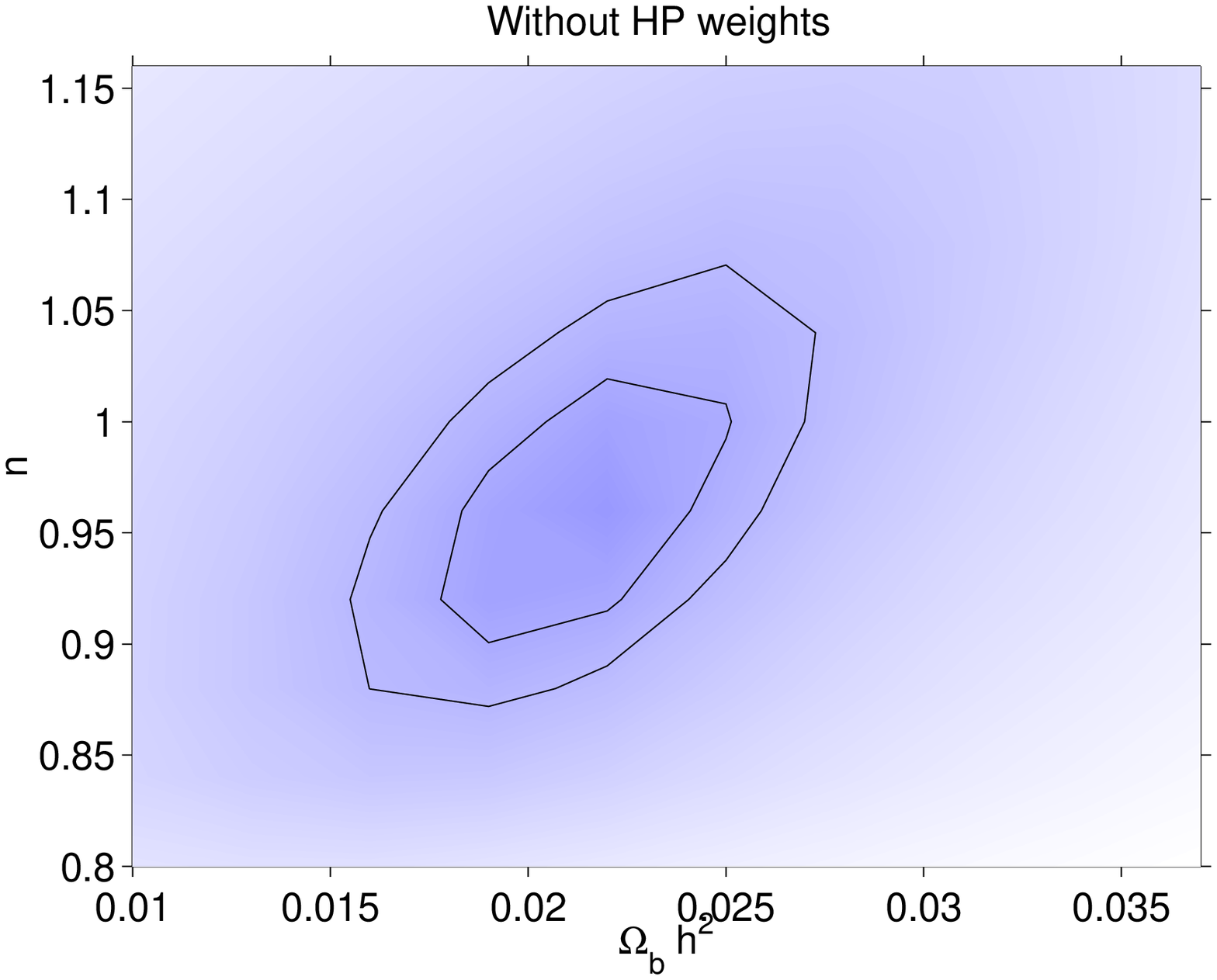,width=7.5cm}
\qquad\qquad
\epsfig{file=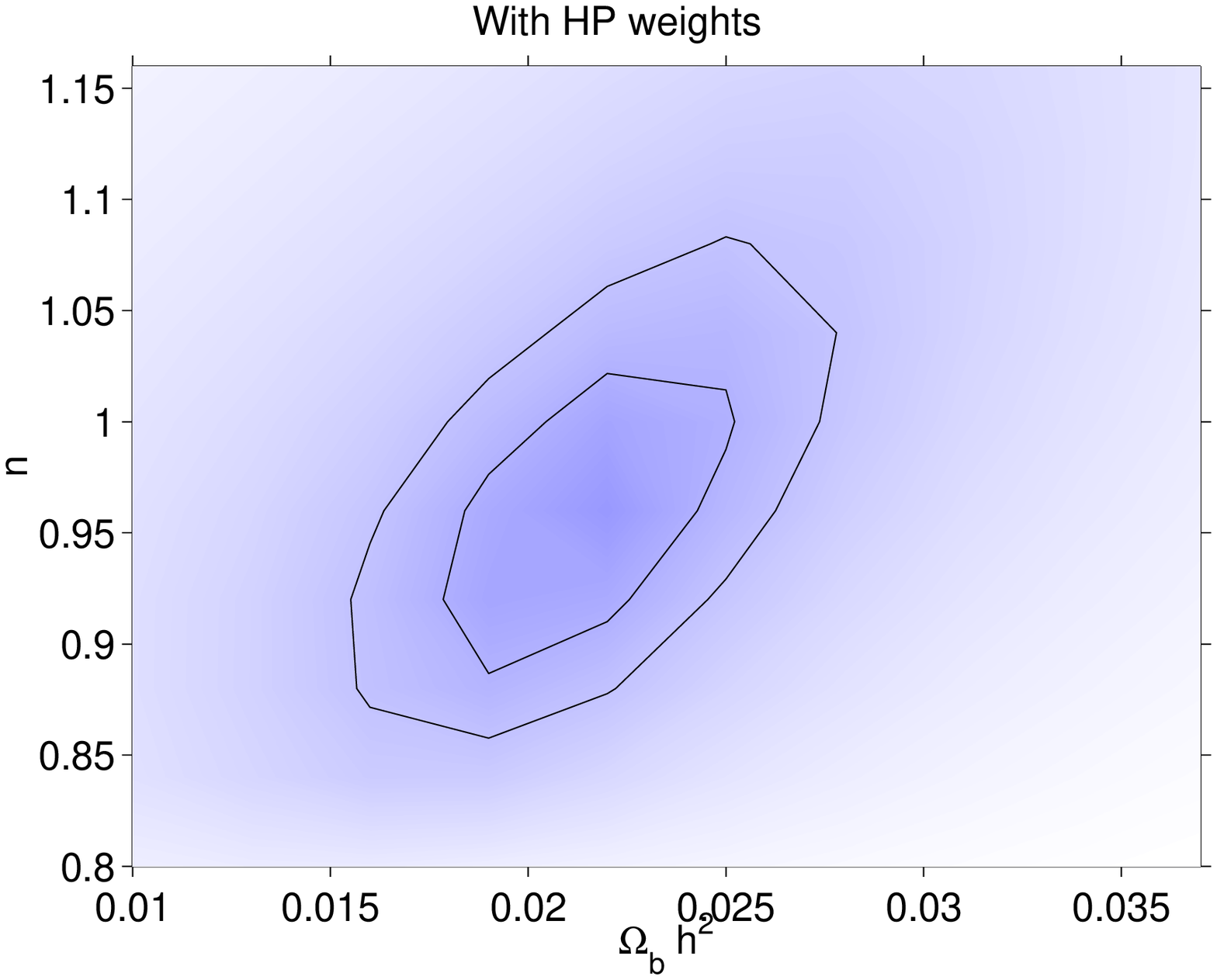,width=7.5cm}
}
\caption{The standard posterior 
$\overline{\Pr}(n,\Omega_bh^2|\vect{D},H_0)$ (left) and the
hyperparameters posterior $\overline{\Pr}(n,\Omega_bh^2|\vect{D},H_1)$
(right) obtained by marginalisation over the
parameters $Q$, $h$ and $\Omega_m$, assuming no calibration or beam
uncertainties in the data plotted in Fig.~\ref{cmbdata}.}
\label{obnresults}
\end{center}
\end{figure*}

Since $\overline{\Pr}(\btheta|\vect{D},H_0)$ 
and $\overline{\Pr}(\btheta|\vect{D},H_1)$ are each calculated over a
hypercube in the 5-dimensional parameter space, one can calculated the
evidence integrals (\ref{h0evid}) and (\ref{h1evid}) directly, and
one finds the exact evidence ratio to be
\[
\frac{\Pr(\vect{D}|H_1)}{\Pr(\vect{D}|H_0)} = 0.05.
\]
This indicates that the data do not support the inclusion of and
subsequent marginalisation over a free hyperparameter 
weight $\alpha_k$ $(k=1,2,3,4)$ for each dataset.
In other words, as expected, the datasets
are statistically consistent both with one another and 
with the range of theoretical models with which they have been compared.

For illustration, let us consider the constraints placed by the current 
CMB data on the parameters $n$ and
$\Omega_b h^2$ with and without the inclusion of hyperparameter weights.
The limits placed on these parameters by the standard
and hyperparameters approaches respectively are easily obtained
by marginalising the corresponding posteriors 
$\overline{\Pr}(\btheta|\vect{D},H_0)$ 
and $\overline{\Pr}(\btheta|\vect{D},H_1)$ over the remaining 
parameters $Q$, $h$ and $\Omega_m$. The resulting distributions
are shown in Fig.~\ref{obnresults}, together with corresponding 
the 68 and 95 per cent confidence contours. 
Although the above evidence ratio shows that the data do 
not favour marginalisation over
hyperparameter weights, we see that 
their inclusion does not significantly affect the
constraints placed on the parameters. This illustrates that, for
statistically consistent datasets, the standard and hyperparameters
techniques may be use interchangeable without degrading the
constraints on cosmological parameters. In particular, we note
that the current CMB data are consistent with the scale invariant 
spectrum $n=1$ and the nucleosynthesis constraints on $\Omega_b h^2$.
Indeed, by performing an additional marginalisation over $n$ or
$\Omega_bh^2$ we obtain (to two significant figures)
the one-dimensional 68-per cent confidence
intervals $0.019 < \Omega_bh^2 < 0.023$ and
$0.92 < n < 1.00$ for {\em both}
the case $H_0$ and $H_1$.

\subsection{Marginalisation over calibration and beam uncertainties}

So far our analysis has ignored the calibration and beam
uncertainties associated with each dataset. Any meaningful analysis
of these data must, however, include these effects of these
`nuisance' parameters. As with hyperparameter weights, the correct
approach is to assign some prior to these parameters and then
marginalise over them. For calibration and beam uncertainties,
one does have a priori knowledge of both the expectation value
of each parameter {\em and} the range of values it might take
(i.e. its variance). Thus, as shown in section~\ref{weight_prior}, it is
appropriate to adopt independent Gaussian priors on the calibration
and beam uncertainties. As discussed in detail by Bridle
et al. (2001), for Gaussian likelihood functions of the form
(\ref{h0likea}), one can again perform this marginalisation analytically.
Moreover, the resulting
likelihood function after marginalisation is also of the form (\ref{h0likea}), 
but with a modified covariance matrix $\vect{V}'_k$. 

Using these modified datasets, one can then
calculate the posteriors $\overline{\Pr}(\btheta|\vect{D},H_0)$ 
and $\overline{\Pr}(\btheta|\vect{D},H_1)$ over the same
hypercube in the 5-dimensional parameter space as used above,
and evaluate the evidence integrals (\ref{h0evid}) and (\ref{h1evid}) 
directly. In this case, the evidence ratio is found to be
\[
\frac{\Pr(\vect{D}|H_1)}{\Pr(\vect{D}|H_0)} = 0.08,
\]
which illustrates that the inclusion of
and marginalisation over hyperparameter weights is once more
unwarranted. The corresponding marginalised posteriors in the
$(n,\Omega_b h^2)$-plane, with and without marginalisation over 
are again almost identical and closely resemble those plotted in
Fig.~\ref{obnresults}. One obtains the one-dimensional 68-per cent
confidence intervals $0.018 < \Omega_b h^2 < 0.024$ and
$0.90 < n < 0.99$ for the case $H_0$ (without hyperparameter weights) 
and $0.018 < \Omega_b h^2 < 0.024$ and
$0.90 < n < 1.00$ for the case $H_1$ (with hyperparameter weights).
Once again, we see that the inclusion of and marginalisation over 
hyperparameter weights has little effect on 
the parameter constraints imposed by the data.

\subsection{Evidence for calibration and beam uncertainties}

Although the evidence ratios given above show that
the current CMB power spectrum data clearly do not require
the use of hyperparameter weights in their analysis,
it is also interesting to compare the relative evidence values
for the cases with and without marginalisation over calibration and
beam uncertainties (for both $H_0$ and $H_1$).
The evidence
values corresponding to each of the 4 cases we have considered
are given in Table~\ref{table1}, where they have been rescaled
so that the evidence is unity for the case where no hyperparameter weights are
introduced and no marginalisation over calibration and beam
uncertainties is performed.
We see that the largest evidence value is obtained for
the case in which no marginalisation is performed 
over hyperparameter weights or calibration and beam uncertainties.
While one might expect from the data plot in Fig.~\ref{cmbdata}
that the introduction of and marginalisation over
hyperparameter weights might not be favoured, it is surprising that
the marginalisation over the calibration and beam uncertainties
also reduces the evidence for the data. This means that, when compared with
the range theoretical models discussed above, the
probabilty of obtaining the observed data is highest if one assumes
{\em no} uncertainty exists in the calibration and beam properties
adopted by each experiment, rather than marginalising over the reported
uncertainties in these parameters. Thus we see that the evaluation of
evidences has provided quantitative support for our earlier naive
observation that the CMB datasets plotted in Fig.~\ref{cmbdata}
agree remarkably well given the 5--20 per cent calibration 
uncertainties alone (which are not plotted). 
We note here simply that this eventuality is unlikely to occur by chance.
\begin{table}
\begin{tabular}{lll}
\hline
            & No HP weights & With HP weights \\
\hline
No cal/beam marginalisation    & $1.0$  & $5\times 10^{-2}$ \\
With cal/beam marginalisation  & $4\times 10^{-4}$ & $3 \times 10^{-5}$ \\
\hline
\end{tabular}
\caption{The relative Bayesian evidences for the obtaining the
current CMB power spectrum data in the 4 cases discussed in the text.}
\label{table1}
\end{table}

\section{Conclusions}
\label{concs}

We have presented a general account of the use of hyperparameters in the
analysis of cosmological datasets. 
In particular, we have concentrated on applying
the hyperparameters technique to the problem of weighting 
different datasets in a joint analysis aimed at estimating
some set of cosmological parameters. The basic approach is to
assign a free hyperparameter weight to each dataset and then
perform a marginalisation over these hyperparameters.
This method allows the statistical
properties of the data themselves to determine the effective weight
given to each dataset and is in sharp contrast to the more 
common practice of excluding certain
datasets from the analysis, which hence are assigned a weight of zero,
and analysing the remaining datasets with equal weights.

Assuming the expected value of the hyperparameter weight on each
dataset to be unity, we find that the prior on each hyperparameter
should be of exponential form, rather than the more 
commonly used improper Jeffrey's
prior. In either case, the
marginalisation over the hyperparameters may be performed analytically
for Gaussian likelihood functions.
Since the exponential prior is properly normalised, in this case one
may also calculate the Bayesian evidence for the data given the
hyperparameters model. This evidence value can then be compared with the
corresponding evidence for the data in the absence of hyperparameters,
in order to determine whether the data warrant the inclusion
of hyperparameters in the first instance. In each case, the evidence
may be calculated either by direct integration, for low-dimensionality 
parameter spaces, or approximated straightforwardly by assuming
each posterior to be Gaussian near its peak. 

The hyperparameter approach to weighting datasets is illustrated by
applying it to the classic toy model of fitting a straight to a number
of datasets. We find the hyperparameters technique correctly infers
the existence of systematic errors
and/or misquoted random errors in the datasets. In such cases, 
the evidence ratio for the hyperparameters and standard approaches
clearly indicates that the data warrant the introduction of the
hyperparameters. Nevertheless, 
in the case where no systematic errors exists, and
the random errors are accurately quoted, the evidence ratio
correctly indicates that hyperparameters should not be included in the
analysis.

Finally, the hyperparameters technique is applied to the latest
measurements of the cosmic microwave background (CMB) power spectrum
by the BOOMERANG, MAXIMA and DASI experiments, together with
the large-scale normalisation of the power spectrum provided by the COBE satellite. The evidence ratio between the hyperparameters and
standard approach shows that 
the current CMB
datasets do not warrant the inclusion of 
hyperparameters weights and a subsequent marginalisation over them.
In other words, the
datasets are statistically consistent both with one another and with
the range of theoretical models with which they were
compared. Nevertheless,
the inclusion of hyperparameters in shown not to lessen the
constraints imposed by the data on cosmological parameters.

The analysis is repeated for the case in which one first marginalises
over the beam and calibration uncertainties associated with each
dataset, and again the evidence ratio shows 
that the inclusion of hyperparameter weights is unwarranted. 
Of more interest is the comparison of evidences with and without
marginalisation over calibration and beam uncertainties. This shows
that the evidence for the data is largest if one assumes no uncertainty
in the calibration and beam properties, but instead adopt
the values 
corresponding to those implied by the published CMB power spectrum
points for each dataset. This is unlikely to occur by chance.

\subsection*{ACKNOWLEDGMENTS}
MPH thanks Steve Gull and Anthony Lasenby 
for some illuminating discussions concerning 
Bayes' theorem and probability theory. The authors also thank
Antony Lewis for providing the $C_\ell$-grid and Anze Slozar
for the use of his $C_\ell$-grid interpolation routines.
SLB acknowledges Selwyn College, Cambridge, for support in the form
of the Trevelyan Research Fellowship.

\bsp 
\label{lastpage}
\end{document}